\documentclass[12pt,a4paper]{article}
\usepackage{a4wide}
\usepackage{latexsym}
\usepackage{epsf}
\usepackage{amssymb}
\makeatletter
\@addtoreset{equation}{section}
\makeatother

\begin{document}

\begin{flushright}
\small
IFT-UAM/CSIC-99-32\\
{\bf hep-th/9910020}\\
October $1$st, $1999$
\normalsize
\end{flushright}

\begin{center}


\vspace{1cm}

{\Large {\bf The General, Duality-Invariant Family of Non-BPS}}\\
\vspace{.5cm}
{\Large {\bf Black-Hole Solutions of  $N=4$, $d=4$ Supergravity}}

\vspace{2cm}


{\bf\large Ernesto Lozano-Tellechea}${}^{\spadesuit}$
\footnote{E-mail: {\tt Ernesto.Lozano@uam.es}}
{\bf\large and Tom\'as Ort\'{\i}n}${}^{\spadesuit\clubsuit}$
\footnote{E-mail: {\tt tomas@leonidas.imaff.csic.es}}

\vspace{1cm}
${}^{\spadesuit}$\ {\it Instituto de F\'{\i}sica Te\'orica, C-XVI,
Universidad Aut\'onoma de Madrid \\
E-28049-Madrid, Spain}

\vskip 0.2cm

${}^{\clubsuit}$\ {\it I.M.A.F.F., C.S.I.C., Calle de Serrano 113 bis\\ 
E-28006-Madrid, Spain}

\vspace{2cm}


{\bf Abstract}

\end{center}

\begin{quotation}

\small

We present the most general family of stationary point-like solutions
of pure $N=4$, $d=4$ Supergravity characterized by {\it completely
  independent} electric and magnetic charges, mass, angular momentum
and NUT charge plus the asymptotic values of the scalars. It includes,
for specific values of the charges all previously known BPS and
non-BPS, extreme and non-extreme black holes and Taub-NUT solutions.

As a family of solutions, it is manifestly invariant under T~and
S~duality transformations and exhibits a structure related to the
underlying special geometry structure of the theory.

Finally, we study briefly the black-hole-type subfamily of metrics and
give explicit expressions for their entropy and temperature.

\end{quotation}

\newpage

\pagestyle{plain}


\section*{Introduction}

The low-energy effective action of the heterotic string compactified
on $T^{6}$ is that of pure $N=4, d=4$ Supergravity coupled to $N=4$
super Yang-Mills. It is possible to truncate consistently this theory
to the simpler pure supergravity theory. From the string theory point
of view the truncation consists in introducing always equal numbers of
Kaluza-Klein and winding modes for each cycle. The truncated theory
still exhibits S~and T~dualities and, thus, pure $N=4,d=4$
Supergravity provides  a simple framework in which to study
classical solutions which still can be considered as solutions of the
full effective String Theory. The bosonic sector of this theory is
also known in the literature as ``Dilaton-Axion Gravity'' or as
``Einstein-Maxwell Dilaton-Axion Theory'' when only a single vector
field is considered.

Perhaps the most interesting solutions of the 4-dimensional string
effective action are the black-hole type ones\footnote{For a review of
  black holes in toroidally compactified string theory see
  e.g.~\cite{kn:Cv1} and \cite{kn:Y}.} since they constitute the best
testing ground for the Quantum Gravity theory contained in String
Theory. It is believed that a good Quantum Gravity theory should be
able to explain in terms of microscopic degrees of freedom the values
of the macroscopic thermodynamical quantities found classically and
semiclassically. There has been some success in this respect for
supersymmetric (``BPS-saturated'') and near-supersymmetric black holes
although the results are to be interpreted carefully since the
supersymmetric limit is singular in many respects.
 
A great deal of effort has been put in finding the most general
families of black-hole solutions whose thermodynamical properties
should exhibit also invariance (or, rather, covariance) under the
duality symmetries of the theory and covering the supersymmetric and
non-supersymmetric cases and, further, covering stationary (not
static) cases.

The first two examples of this kind of families of solutions were
found in Ref.~\cite{kn:KO}\footnote{In the much simpler context of
  pure $N=2,d=4$ Supergravity the IWP solutions of
  Refs.~\cite{kn:Pe,kn:IW,kn:To} also have this property.}.  The
first family of solutions corresponds to non-supersymmetric, static
black-hole solutions and the second to supersymmetric, static,
multi-black-hole solutions of $N=4,d=4$ supergravity. Under the
dualities of the theory, solutions of each family transform into other
solutions of the same family, with the same functional form. Thus,
only the values of the charges and moduli transform. The
supersymmetric solutions are given in terms of two constrained complex
harmonic functions.

Different extensions and properties of these solutions in the context
of Dilaton-Axion Gravity were later obtained in
Refs.~\cite{kn:GaGaK2,kn:GaK,kn:Rog,kn:GaGaK,kn:Rog2,kn:GaK2,kn:CG,kn:Ba,
  kn:GaL,kn:GaS,kn:Rog3}.

A main step forward was given in Ref.~\cite{kn:FKS} where it was
realized that the form of the above supersymmetric solutions was
dictated the special geometry of the associated $N=2,d=4$ Supergravity
theory.  The two complex harmonic functions are associated to
coordinates and certain components of the metric are associated to the
K\"ahler potential and the holomorphic vector.  It was found that
similar Ansatzs could be used in other $N=2,d=4$ Supergravity theories
with different matter multiplets and K\"ahler potentials.

Finally, in Refs.~\cite{kn:To2,kn:BKO3} the most general
supersymmetric black-hole-type solutions of pure $N=4,d=4$
Supergravity (SWIP solutions) where found. The only difference with
those of Ref.~\cite{kn:KO} is that the complex harmonic functions are
now completely arbitrary and unconstrained. This automatically allows
for the introduction of angular momentum and NUT charge in the
solutions.  In fact the constraint simply meant that these charges
were not allowed. The generating solution for regular, supersymemtric,
$N=8$ supergravity black hole solutions has been found in
Ref.~\cite{kn:BFT}.

Similar supersymmetric solutions were later found for other $N=2,d=4$
Supergravity theories \cite{kn:BLS} with vector
multiplets\footnote{For a review on supersymmetric black hole
  solutions of supergravity theories see e.g.~Ref.~\cite{kn:DAF}.}.

For the non-supersymmetric solutions of Ref.~\cite{kn:KO} the story
has been different since no clear relation with the underlying special
geometry was established. In Ref.~\cite{kn:KW} a general recipe for
obtaining non-supersymmetric solutions from supersymmetric solutions
in $N=2,d=4$ Supergravity theories, previously used in other contexts,
was shown to work for {\it static} black holes: one simply has to
deform the metric with the introduction of a non-supersymmetry
(non-extremality) function.

What has to be done in more general cases (stationary, for instance) is
far from clear and general duality-invariant families of stationary
non-supersymmetric solutions are not available in the literature and
no recipe to build them is known.

In this paper we present such a general duality-invariant family of
stationary non-supersymmetric solutions of pure $N=4,d=4$ Supergravity
characterized by completely independent electric and magnetic charges,
mass, angular momentum and NUT charge plus the asymptotic values of
the scalar fields\footnote{When we talk about general solutions we are
  implicitly excluding the possibility of having primary scalar hair.
  Solutions with primary scalar hair are in all known cases (see,
  e.g.~\cite{kn:ALC2}), singular (providing evidence for the never
  proven ``no-hair theorem'') and, being interested in true black
  holes with event horizons covering all the physical singularities,
  these cases are not important for us and in the solutions which we
  are going to present the scalar charges are always completely
  determined by the $U(1)$ charges. Nevertheless, it should be pointed
  out that more general solutions (some of them supersymmetric) which
  include primary scalar hair must exist and should be related to the
  ones given here by formal T~duality in the time direction
  \cite{kn:AMO}.}.

The rest of the paper is organized as follows: in
Section~\ref{sec-n4d4} we describe the bosonic sector of $N=4,d=4$
Supergravity theory. In Section~\ref{sec-gensolution} we give and
study the general family of solutions we relate it to others already
known.  In Section~\ref{sec-bhsolution}, we focus our attention in the
black hole type subfamily of metrics and calculate the explicit values
for their entropy and the temperature, showing that also these
quantities can be put in a manifestly duality- invariant form.
Section~\ref{sec-conclusions} contains our conclusions. The Appendices
contain the definitions of the different charges we use and their
duality-invariant combinations.


\section{$N=4$, $d=4$ Supergravity}
\label{sec-n4d4}


\subsection{Description of the System and Equations of Motion}

The bosonic sector of pure $N=4,d=4$ Supergravity contains two real
scalar fields (axion $a$ and dilaton $\phi$), six Abelian vector
fields $A_{\mu}^{(n)}$ (which we generalize to an arbitrary number
$N$) and the metric $g_{\mu\nu}$. The action reads\footnote{Our
  conventions coincide with those of Ref.~\cite{kn:MO}. In particular,
  we use mostly minus signature and Hodge duals are defined such that
  $^{\star}F^{(n)\, \mu\nu}=\frac{1}{2\sqrt{|g|}}
  \epsilon^{\mu\nu\rho\sigma} F_{\rho\sigma}^{(n)}$ with
  $\epsilon^{0123}=+1$.}

\begin{equation}
\label{Action}
S  =  {\textstyle\frac{1}{16\pi}} \int d^{4}x \sqrt{|g|}\, \left\{ R 
+2(\partial\phi)^{2} + {\textstyle\frac{1}{2}}e^{4\phi}(\partial a)^{2}
-e^{-2\phi}\sum_{n=1}^{N} F^{(n)}F^{(n)} 
+a \sum_{n=1}^{N} F^{(n)}\ ^{\star}F^{(n)} \right\}\, .
\end{equation}

The axion and dilaton are combined into a single complex scalar field,
the {\it axidilaton} $\lambda$:

\begin{equation}
\lambda=a+ie^{-2\phi}\, .  
\end{equation}

For each vector field we can also define its $SL(2,\mathbb{R})$-dual,
which with our conventions will be given by:

\begin{equation}
\label{SL2R-Duals}
\tilde{F}^{(n)}{}_{\mu\nu} \equiv e^{-2\phi}\ ^{\star}F^{(n)}{}_{\mu\nu}
+aF^{(n)}{}_{\mu\nu}\, .
\end{equation} 

\noindent The equations of motion derived from the
action~(\ref{Action}) plus the Bianchi identities for the vector
fields can be written as follows:

\begin{eqnarray} 
\nabla_{\mu} {}^{\star}\tilde{F}^{(n)\, \mu\nu} & = & 0 \, , 
\label{Maxwell}\\
& & \nonumber \\
\nabla_{\mu} {}^{\star}F^{(n)\,\mu\nu} & = & 0\, , \\
& & \nonumber \\
\nabla^{2}\phi-{\textstyle\frac{1}{2}}e^{4\phi}(\partial a)^{2}-
{\textstyle\frac{1}{2}}e^{-2\phi}\sum_{n=1}^{N}F^{(n)}F^{(n)}
& = & 0 \, ,\\
& & \nonumber \\
\nabla^{2}a+4\partial_{\mu}\phi\, \partial^{\mu}a
-e^{-4\phi}\sum_{n=1}^{N}F^{(n)}\ ^{\star}F^{(n)}
& = & 0\, , \\
& & \nonumber \\
R_{\mu\nu}+2\partial_{\mu}\phi\partial_{\nu}\phi+
{\textstyle\frac{1}{2}} e^{4\phi}\partial_{\mu}a\partial_{\nu}a
 -2e^{-2\phi}  \sum_{n=1}^{N}\left(F^{(n)}{}_{\mu\rho}F^{(n)}{}_{\nu}{}^{\rho}
-{\textstyle\frac{1}{4}}g_{\mu\nu}F^{(n)}F^{(n)}\right) 
& = & 0\, .
\end{eqnarray}

Observe that we have written the Maxwell equations as the Bianchi
equations for the $SL(2,\mathbb{R})$ duals. Therefore $N$ dual vector
potentials $\tilde{A}_{\mu}^{(n)}$ defined by

\begin{equation}
\tilde{F}_{\mu\nu}^{(n)}=\partial_{\mu}\tilde{A}_{\nu}^{(n)}-
\partial_{\nu}\tilde{A}_{\mu}^{(n)}\, ,
\end{equation}

\noindent exist locally.

The axidilaton parametrizes an $SL(2,\mathbb{R})/SO(2)$
coset~\cite{kn:JHS2}, the equations of motion being invariant under
global $SL(2,\mathbb{R})$ (``S~duality'') transformations.  If
$\Lambda$ is an $SL(2,\mathbb{R})$ matrix
 
\begin{equation}
\Lambda = \left( \begin{array}{cc} 
             a & b \\ 
             c & d \\
             \end{array}\right)\, , 
\hspace{1cm}
ad-bc=1\, ,
\end{equation}

\noindent then the vector fields and their duals transform as 
doublets

\begin{equation}
\left( 
\begin{array}{c} 
\tilde{F}^{(n)}{}_{\mu\nu} \\
\\
F^{(n)}{}_{\mu\nu} \\ 
\end{array} 
\right) 
\longrightarrow 
\Lambda
\left( 
\begin{array}{c} 
\tilde{F}^{(n)}{}_{\mu\nu} \\
\\
F^{(n)}{}_{\mu\nu} \\ 
\end{array} 
\right) \, ,
\label{Transformation-A}
\end{equation}

\noindent and the axidilaton transforms according to

\begin{equation}
\label{Transformation-l}
\lambda \longrightarrow
\frac{a\lambda+b}{c\lambda+d}\, .
\end{equation}

This is an electromagnetic duality rotation that acts on the dilaton.
From the point of view of String Theory, this is the 4-dimensional
string coupling constant. Hence the name S~duality.

Furthermore the $N$ (6 in the SUGRA theory) vector fields can be
$SO(N)$-rotated. These are ``T~duality'' transformations (perturbative
from the String Theory point of view). The full duality group is, then
$SL(2,\mathbb{R})\otimes SO(N)$.


\section{The General Solution}
\label{sec-gensolution}

We now present the family of solutions.  All the fields in our
solutions may be expressed in terms of two {\em fixed} complex
harmonic functions of the three dimensional Euclidean space, ${\cal
  H}_1$ and ${\cal H}_2$, a set of $N$ complex constants $k^{(n)}$, a
``non-extremality'' function $W$ and a background 3-dimensional metric
${}^{(3)}\gamma_{ij}$. In all of them appear the physical constants
defined in Appendix~\ref{sec-parameters}. Only $\Upsilon$, the
axidilaton charge, is not independent. The harmonic functions are

\begin{equation}
\label{Hs}
\left\{
\begin{array}{rcl}
{\cal H}_{1} & = & 
\frac{1}{\sqrt{2}}
e^{\phi_{0}}e^{i\beta} 
\left( \lambda_{0} 
+{\displaystyle \frac{\lambda_{0}{\mathfrak{M}}+\bar{\lambda}_{0}\Upsilon}
{\tilde{\rho}}}
\right)\, ,\\
 & & \\ 
{\cal H}_{2} & = & 
\frac{1}{\sqrt{2}}e^{\phi_{0}}e^{i\beta} 
\left(1+{\displaystyle\frac{{\mathfrak{M}} 
+\Upsilon}{\tilde{\rho}}}\right)\, ,\\
\end{array}
\right.
\end{equation}

\noindent where $\tilde{\rho}^{2} \equiv x^{2}+y^{2}+(z+i\alpha)^{2}$ is the
usual complex radial coordinate, and $\beta$ is an arbitrary,
unphysical real number related to the duality transformation of these
functions under $SL(2,\mathbb{R})$ (see the explanation in
Section~\ref{sec-dualproperties}). The complex constants are

\begin{equation}
\label{ks}
k^{(n)}=-{\textstyle\frac{1}{\sqrt{2}}}e^{-i\beta}
\frac{ {\mathfrak{M}}\Gamma^{(n)}+ \overline{\Upsilon \Gamma^{(n)}} }
{ |{\mathfrak{M}}|^{2} - |\Upsilon|^{2} }\, .
\end{equation} 

In supersymmetric cases (e.g.~Ref.~\cite{kn:BKO3}) it is useful to
introduce oblate spheroidal coordinates which are related to the
ordinary Cartesian ones by:

\begin{equation} 
\label{Spheroidal}
\left\{
\begin{array}{ccl}
x & = & \sqrt{r^{2}+\alpha^{2}}\ \sin{\theta} \cos{\varphi}\, , \\
  &   & \\
y & = & \sqrt{r^{2}+\alpha^{2}}\ \sin{\theta} \sin{\varphi}\, , \\  
  &   & \\
z & = & r \cos{\theta}\, . \\
\end{array}
\right. 
\end{equation}

\noindent The three dimensional Euclidean metric is written in these
coordinates in the following way:

\begin{equation}
\label{Flat-Sph}
d\vec{x}^{\, 2} =
{\displaystyle\frac{r^{2}+\alpha^{2}\cos^{2}\theta}{r^{2}+\alpha^{2}}}\,  
dr^{2}+  
\left( r^{2}+\alpha^{2}\cos^{2}\theta \right) \ d\theta^{2}
+ \left( r^{2}+\alpha^{2} \right)\sin^{2}\theta \  d\varphi^{2}\, .
\end{equation}

In terms of~(\ref{Spheroidal}) the radial coordinate $\tilde{\rho}$
that appears in (\ref{Hs}) may be expressed as
$\tilde{\rho}=r+i\alpha\cos\theta$. Furthermore, in these new
coordinates, the ``non-extremality'' function has a simple form:

\begin{equation}
W=1-\frac{r_{0}^{\ 2}}{r^2+\alpha^{2}\cos^{2}\theta}\, , 
\end{equation}

\noindent where $r_{0}$, given by

\begin{equation}
\label{r0}
r_{0}^{\, 2}= |{\mathfrak{M}}|^{2}+|\Upsilon|^{2}
-\sum_{n=1}^{N}|\Gamma^{(n)}|^{2}\, .
\end{equation}

\noindent  is usually called ``extremality parameter''
in the static cases. In stationary cases, though, $r_{0}=0$ means that
the solution is supersymmetric but in general it is not an extreme
black hole (nor a black hole). Thus, a more appropriate name is {\it
  supersymmetry parameter}. The {\it extremality parameter} will be
$R_{0}^{2}= r_{0}^{2}-\alpha^{2}$.

Finally, the last ingredient is the background metric ${}^{(3)}\gamma_{ij}$

\begin{equation}
\label{Non-Flat}
\begin{array}{rcl}
d\vec{x}^{\, 2} & = & {}^{(3)}\gamma_{ij}dx^{i}dx^{j} \\
& & \\
& &  
\hspace{-1.7cm}
={\displaystyle\frac{r^{2}+\alpha^{2}\cos^{2}\theta-r_{0}^{2}}
{r^{2}+\alpha^{2}-r_{0}^{2}}}\, dr^{2}+  
\left( r^{2}+\alpha^{2}\cos^{2}\theta-r_{0}^{\ 2} \right) d\theta^{2}+
+\left( r^{2}+\alpha^{2}-r_{0}^{\ 2} \right)\sin^{2}\theta d\varphi^{2}\, ,
\end{array}
\end{equation}

\noindent which differs from (\ref{Flat-Sph}) in non-supersymmetric 
($r_{0}\neq 0$) cases and is not flat\footnote{In~(\ref{eq:solution})
  as well as in~(\ref{Non-Flat}) the $x^{i}$ label the coordinates
  $r$, $\theta$ and $\varphi$ for $i=1,2,3$ respectively.}.  This is
an important qualitative difference between the usual supersymmetric
IWP-type \cite{kn:Pe,kn:IW} metrics (e.g.~those of
Refs.~\cite{kn:To,kn:To2,kn:BKO3}) and our solution.

We can now describe the solutions. They take the form

\begin{equation}
\left\{
\label{eq:solution}
\begin{array}{rcl}
ds^{2} & = & e^{2U}W\left(dt+\omega_{\varphi} d\varphi\right)^{2}-
     e^{-2U}W^{-1}\ ^{^{(3)}}\gamma_{ij}dx^{i}dx^{j}\, ,  \\
& & \\
A^{(n)}{}_{t} & = & 2e^{2U}\, 
\Re{\rm e}\left( k^{(n)}{\cal H}_2\right)\, , \\
& & \\
\tilde{A}^{(n)}{}_{t} & = & 2e^{2U}\, 
\Re{\rm e}\left( k^{(n)}{\cal H}_1\right)\, , \\
& & \\
\lambda & = & \displaystyle{\frac{{\cal H}_1}{{\cal H}_2}}\, , 
\end{array}
\right.
\end{equation}

\noindent where

\begin{equation}
e^{-2U}=2\ \Im{\rm m}\left( {\cal H}_{1} \bar{\cal H}_{2} \right)=
1+2 \Re{\rm e} \left( \frac{{\mathfrak{M}}}{r+i\alpha \cos{\theta}} \right)+
 \frac{|{\mathfrak{M}}|^{2}-|\Upsilon|^{2}}{r^2+\alpha^{2}\cos^{2}\theta}\, ,
\end{equation} 

\noindent and where

\begin{equation}
\begin{array}{rcl}
\omega_{\varphi}  & = & 
2n\cos{\theta} +\alpha\sin^{2}\theta\left( e^{-2U} W^{-1} -1\right) \\
& & \\
& = & 
{\displaystyle\frac{2}{r^2+\alpha^{2}\cos^{2}\theta-r_{0}^{2}}}\, 
\times\\
& & \\
& &  
\times
\left\{ n\cos{\theta} \left(r^{2}+\alpha^{2}-r_{0}^{2} \right)
+\alpha \sin^2\theta 
\left[ mr+ \frac{1}{2}       
\left( r_{0}^{\ 2}+|{\mathfrak{M}}|^2-|\Upsilon|^2\right) 
\right]\right\}\, .
\end{array}
\end{equation}

\noindent $\omega_{\varphi}$ can be interpreted as the unique 
non-vanishing covariant component of a 1-form $\omega=\omega_{i}
dx^{i}$ which for $\alpha\neq 0$ is a solution of the following
equation in 3-dimensional space:

\begin{equation}
\label{Omega-H}
{}^{\star}d\omega  -e^{-2U} {}^{\star} d\mu  
-W^{-1} \Re{\rm e}\left( {\cal H}_{1}d\bar{\cal H}_{2} 
-\bar{\cal H}_{2}d{\cal H}_{1}\right) = 0\, ,
\end{equation}

\noindent where $\mu$ is also a 1-form whose only non-vanishing
component $\mu_{\varphi}$ is given by

\begin{equation}
\mu_{\varphi} = \frac{r_{0}^{2}}{\alpha}\, 
\frac{r^{2}+\alpha^{2}-r_{0}^{2}}
{r^{2}+\alpha^{2}\cos^{2}\theta-r_{0}^{2}}\, , 
\end{equation}

\noindent and where the 3-dimensional background metric 
${}^{(3)}\gamma_{ij}$ has to be used in the Hodge duals.

For $\alpha=0$ the $\mu$ term in Eq.~(\ref{Omega-H}) has to be
eliminated and the equation tales the form

\begin{equation}
\label{Omega-Halphazero}
{}^{\star}d\omega  
-W^{-1} \Re{\rm e}\left( {\cal H}_{1}d\bar{\cal H}_{2} 
-\bar{\cal H}_{2}d{\cal H}_{1}\right) = 0\, .
\end{equation}

We can also write our solution in the standard form used to describe
general rotating black holes, which will be useful to describe the
structure of the singularities:

\begin{eqnarray}
ds^{2} & = & \frac{\Delta-\alpha^2\sin^2\theta}{\Sigma}\, dt^2+
            2\alpha\sin^2\theta\frac
            {\Sigma+\alpha^2\sin^2\theta-\Delta}{\Sigma}
            \, dt d\varphi- \nonumber\\
& & \nonumber \\              
& & -\frac{\Sigma}{\Delta}\, dr^2-\Sigma\, d\theta^2
              -\frac{\left(\Sigma+\alpha^2\sin^2\theta \right)^2-
                     \Delta\alpha^2\sin^2\theta}{\Sigma}
               \sin^2\theta\, d\varphi^2\, ,
\label{Kerr-Type}
\end{eqnarray}

\noindent where

\begin{equation}
\begin{array}{rcl}
\Delta & = & r^{2} -R_{0}{}^{2} = r^{2}+\alpha^{2}-r_{0}{}^{2}\, , \\
& & \\
\Sigma & = & (r+m)^2+(n+\alpha\cos\theta)^2-|\Upsilon|^2\, .
\end{array}
\label{Delta-Sigma}
\end{equation}

This completes the description of the general solution. Now we are
going to describe its properties.


\subsection{Duality Properties}
\label{sec-dualproperties}

We can study the effect of duality transformations in two ways which
are fully equivalent in this family of solutions: we can study the
effect of the transformations of the fields or simply the effect of
the transformations on the physical constants. One of the main
features of our family of solutions is precisely this equivalence: we
can simply transform the physical constants (adding ``primes'')
because the functional form of the solutions always will remain
invariant.

Let us, then, study the effect of $SL(2,\mathbb{R})$ transformations
of the charges $\mathfrak{M},\Gamma^{(n)},\Upsilon$ and moduli
$\lambda_{0}$ in Appendix~\ref{sec-parameters}.

Both the complex harmonic functions ${\cal H}_{1,2}$ and the complex
constants $k^{(n)}$ are defined up to a phase: if we multiply ${\cal
  H}_{1,2}$ by a constant phase and the $k^{(n)}$'s by the opposite
one, the solution remains unchanged. We have made this fact explicit
by including the arbitrary angle $\beta$ in their definition.

As it can take any value, in particular we can require it to change in
the following way when performing an $SL(2,\mathbb{R})$ rotation:

\begin{equation}
e^{i\beta} \longrightarrow 
e^{i\arg(c\lambda_{0}+d)}e^{i\beta}\, .
\end{equation}

\noindent With this choice the  $k^{(n)}$'s are
left invariant, while the pair ${\cal H}_{1,2}$ transforms as a
doublet:

\begin{equation}
\label{Transformation-H}
\left( \begin{array}{c} {\cal H}_{1}^{\prime} \\ 
{\cal H}_{2}^{\prime} \end{array} \right)
=\Lambda
\left( \begin{array}{c} {\cal H}_{1} \\ {\cal H}_{2} \end{array} \right)\, .
\end{equation}

${\cal H}_{1,2}$ appear only through two invariant combinations:
$e^{-2U}$ and $\Re{\rm e}\left( {\cal H}_{1}d\bar{\cal H}_{2}
  -\bar{\cal H}_{2}d{\cal H}_{1}\right)$. The remaining building
blocks of the solution are $\mu$ and $W$ which are invariant if the
supersymmetry parameter $r_{0}$ is invariant. This (first proven in
Ref.~\cite{kn:O1}) is shown in Appendix~\ref{sec-centralcharge}.

Under $SO(N)$ the $k^{(n)}$'s transform as vectors, as they should,
and everything else is invariant.

The relation of the form of these solutions to $N=2$ special geometry
is the same as in the supersymmetric case \cite{kn:FKS,kn:BKO3} and we
will not repeat here that discussion. The only difference is the
introduction of the background metric ${}^{(3)}\gamma$, and the
functions $\mu,W$ which ``deform'' the supersymmetric solution but
have no special meaning from the special geometry point of view.


\subsection{Reduction to Other Known Solutions}

We can now relate our solution to those less general found in the
literature. We can consider two types of solutions: supersymmetric and
non-supersymmetric. The most general family of supersymmetric
black-hole type solutions of $N=4,d=4$ Supergravity (SWIP solutions)
was found in Refs.~\cite{kn:To2,kn:BKO3}. Due to the existence of
supersymmetry, the family contains two arbitrary harmonic functions.
To describe point-like solutions one chooses harmonic functions with a
single pole. Our solutions reduce precisely to these when the
supersymmetry parameter vanishes: $r_{0}=0$. As shown in
Appendix~\ref{sec-centralcharge} in the $r_{0}=0$ limit at least one
of the two possible Bogomol'nyi bounds of $N=4$ Supergravity are
saturated.  In this case $W=1,\mu_{\varphi}=0$, ${}^{(3)}\gamma$
becomes flat (in spheroidal coordinates) and we recover the structure
of the SWIP solutions. The SWIP solutions always saturate one bound
due to the constraints that the constants $k^{(n)}$ satisfy

\begin{equation}
\begin{array}{rcl}
\displaystyle{\sum_{n=1}^{N}}(k^{(n)})^{2} & = & 0\, , \\
& & \\
\displaystyle{\sum_{n=1}^{N}}|k^{(n)}|^{2} & = & \frac{1}{2}\, . 
\end{array}
\label{SWIPk}
\end{equation}     

\noindent while, in our case there is no constraint on the charges (apart
from the one on the scalar charge, associated to the non-hair theorem)

\begin{equation}
\begin{array}{rcl}
\displaystyle{\sum_{n=1}^{N}}(k^{(n)})^{2} & = &
\displaystyle{\frac{-{\mathfrak{M}}\overline{\Upsilon}} 
{\left(|{\mathfrak{M}}|^2-|\Upsilon|^{2}\right)^{2}}\  r_{0}^{\ 2}}\, ,\\
& & \\ 
\displaystyle{\sum_{n=1}^{N}}|k^{(n)}|^{2} & = &
\frac{1}{2}\displaystyle{ \left(1-\frac{|{\mathfrak{M}}|^2+|\Upsilon|^2}
{\left(|{\mathfrak{M}}|^2-|\Upsilon|^2\right)^2}\  r_{0}^{\ 2}
\right)}\, .
\end{array}
\end{equation}    

\noindent In Ref.~\cite{kn:BKO3} it is shown how this solution 
reduces to supersymmetric solutions with angular momentum, NUT charge
etc. Only some of the static ones (those with $1/4$ of the
supersymmetries unbroken) are black holes with a regular horizon.
These include extreme Reissner-Nordstr\"om black holes and their
axion-dilaton generalizations
\cite{kn:G,kn:GHS,kn:STW,kn:KLOPP,kn:O1}. The rest have naked
singularities.

As for the non-supersymmetric solutions, the non-extreme Taub-NUT
axion-dilaton solutions of Ref.~\cite{kn:KKOT} are clearly covered by
our general solution. Further, in Ref.~\cite{kn:GaK} were found
general point-like solutions for a theory with only one vector field
(``axion-dilaton gravity''). We can see that our solutions reduce to
these ones by setting $N=1$. The principal difference is that, in this
particular case, we can fit the analogous of expression~(\ref{Upsilon})
in the definition~(\ref{r0}) for the extremality parameter, giving

\begin{equation}
r_{0,N=1}^{\ 2}=\left(|{\mathfrak{M}}|-|\Upsilon|\right)^2\, ,
\end{equation}

\noindent and inserting this into the metric~(\ref{eq:solution}) we
get exactly Eqs.~(31-35) of Ref.~\cite{kn:GaK} up to a shift in the
radial coordinate.  Although one can immediately see that the
functional form of the metric~(\ref{eq:solution}) does not change very
much from that found in~\cite{kn:GaK}, somewhat different results
appear when considering multiple vector fields, due to the constraints
that the physical parameters obey when only one vector field is
present. This analysis was already done in detail in~\cite{kn:BKO3},
and we refer to this paper for further discussion.

A generalization of the non-extreme solutions of~\cite{kn:GaK} for the
same theory, but with an arbitrary number of vector fields ({\em
  i.e.}, the same theory we are treating), was found in~\cite{kn:GaL}.
However, the solutions reported there concern only the {\em static}
case, and therefore the total number of independent physical
parameters is $2N+4$. As it was shown in that paper, the metric for
the static case is of the ``Reissner-Nordstr\"om-type'', but with a
variable mass factor. It can be seen that, taking the static
($\alpha=0$) limit of our metric~(\ref{eq:solution}), and shifting the
radial coordinate by a quantity $m+\sqrt{|\Upsilon|^2-n^2}$, we
recover the same solution of~\cite{kn:GaL} (Eq.~(7.6) of that
reference) up to redefinitions in the different constants
parametrizing the solution.

Finally, we observe that setting the axidilaton charge equal to zero
(which can be done with appropriate combinations of electric and
magnetic charges) in Eqs.~(\ref{Delta-Sigma}-\ref{Kerr-Type}), we
recover the Kerr-Newman solution in Boyer-Lindquist coordinates (but
with a constant shift equal to the mass in the radial coordinate.
See, {\it e.g.}, Refs.\cite{kn:Tow5,kn:W}).


\section{Black-Hole-Type Solutions}
\label{sec-bhsolution}


\subsection{Singularities}

We now carry out the analysis of the structure of our solutions.
First, we proceed to study the different types of singularities of the
metric. Due to the standard form of $g_{\mu\nu}$ in terms of $\Delta$
and $\Sigma$, the singularities in terms of these functions are those
of all Kerr-type metrics, {\it i.e.}, we have coordinate singularities
at

\begin{equation}
\Delta=0\, , \ \ \ \ \ \theta=0\, , 
\label{Coord-Sing}
\end{equation}

\noindent and a curvature singularity at

\begin{equation}
\Sigma=0\, .
\end{equation}

The first of Eqs.~(\ref{Coord-Sing}) gives the possible horizons. 
To study the different cases, let us shift the radial coordinate 
to recover the Boyer-Lindquist coordinates in which this kind of 
solutions are usually given. If we perform the following rescaling:

\begin{equation}
    r\longrightarrow r-m \, \nonumber
\end{equation}

\noindent then $\Delta$ and $\Sigma$ of~(\ref{Delta-Sigma}) become

\begin{equation}
\begin{array}{rcl}
\Delta & = & (r-m)^2-R_{0}{}^{2}\, , \\
 & & \\
\Sigma & = & r^{2}+\alpha^{2}\cos^{2}\theta-|\Upsilon|^{2}\, ,\\
\end{array}
\end{equation}

\noindent where we also have made the NUT charge $n$ equal to zero 
in order to obtain black-hole-type solutions. In studying the
singularities given by $\Delta=0$ we have three cases to consider:

\begin{flushleft}
${\rm\bf a)}\ \ R_{0}^{2}<0\, ,\,\,\,\, (r_{0}{}^{2} < \alpha^{2})\, .$
\end{flushleft}

Here $\Delta=0$ has no real solutions, we have a naked singularity at
$\Sigma=0$ and no true black hole interpretation is possible. This is
the case of supersymmetric ($r_{0}=0$) rotating ($\alpha\neq 0$)
``black holes''.

\begin{flushleft}
${\rm\bf  b)}\ \  R_{0}^{2}>0\, ,\,\,\,\, (r_{0}{}^{2} > \alpha^{2})\, .$
\end{flushleft}

In this case we have two horizons placed at

\begin{equation}
r_{\pm}= m \pm R_{0}\, .
\end{equation}

\noindent To see if  in this case we have a true black hole we must
verify that the singularity is always hidden by the event horizon. The
region where the singularity is placed is given by the following
equation:

\begin{equation}
\Sigma = 0 \ \Leftrightarrow \ 
r_{\rm sing}^{2} =|\Upsilon|^{2}-\alpha^{2}\cos^{2}\theta\, .
\end{equation}

This is not the usual ``ring singularity'', but a more complicated
2-dimensional {\it surface} in general. Depending on the values of the
charges, this can have the topology of the surface of a torus (maybe
degenerate in certain cases to the surfaces of two concentric
ellipsoids). Whatever its shape is, it is always confined in the region

\begin{equation}
r_{\rm sing}^{2} \leq |\Upsilon|^{2}\, ,
  \label{Sing-Bound}
\end{equation}

\noindent while, on the other hand, the would-be event horizon 

\begin{equation}
r_{+} =m + R_{0} >m \, ,
\end{equation}

\noindent and it will cover the singularity if $m>|\Upsilon|$.
Using the value of $|\Upsilon|$ in terms of the other charges
it is easy to prove

\begin{equation}
\left( |\mathfrak{M}| -|\Upsilon|\right)^{2} > \alpha^{2}\, .
\end{equation}

We can now distinguish two cases:

\begin{flushleft}
${\rm\bf i)}\,\,\,\, |\mathfrak{M}| - |\Upsilon| > |\alpha|\, .$
\end{flushleft}

In this case (setting $n=0$) the horizon covers the singularity and
the object is a true black hole.  Using the expressions in
Appendix~\ref{sec-centralcharge} it is possible to prove that this
happens when both

\begin{equation}
|\mathfrak{M}| > |{\cal Z}_{1,2}|\, ,
\end{equation}

\noindent which is the case allowed by supersymmetry (but not supersymmetric).

\begin{flushleft}
  ${\rm\bf ii)}\,\,\,\, |\mathfrak{M}| - |\Upsilon| < |\alpha|\, .$
\end{flushleft}

In this case there are naked singularities. This is the case  
forbidden by supersymmetry since one can show that in it 
both Bogomol'nyi bounds are simultaneously violated

\begin{equation}
|\mathfrak{M}| < |{\cal Z}_{1,2}|\, .
\end{equation}

\begin{flushleft}
${\rm\bf  c)}\ \  R_{0}^{2}=0\, ,\,\,\,\, (r_{0}{}^{2} =\alpha^{2})\, .$
\end{flushleft}

This is the extremal case, and here we have a single would-be horizon
placed at

\begin{equation} 
\label{Ext-Horizon}
r_{\pm}=m\, .
\end{equation} 

Again, we can distinguish two cases

\begin{flushleft}
  ${\rm\bf i)}\,\,\,\, |\mathfrak{M}| - |\Upsilon| > |\alpha|\,
  ,\,\,\,\,|\mathfrak{M}| > |{\cal Z}_{1,2}|\, .$
\end{flushleft}

In this case the singularity is inside the horizon and we have a true
extreme rotating black hole. This is the case allowed by supersymmetry
(not supersymmetric unless $\alpha=0$).

\begin{flushleft}
  ${\rm\bf i)}\,\,\,\, |\mathfrak{M}| - |\Upsilon| < |\alpha|\,
  ,\,\,\,\,|\mathfrak{M}| < |{\cal Z}_{1,2}|\, .$
\end{flushleft}

The singularity is outside the ``horizon'' and this is not a black
hole.

    
\subsection{Entropy and temperature}

We can now calculate the physical quantities associated to the true
black-hole-type solutions. The entropy of the BH can be worked out by a
straightforward computation of the area of the event horizon. This
gives the following result:

\begin{equation}
    A_{horizon}=4\pi\left(r_{+}^{\ 2}+\alpha^2-|\Upsilon|^2\right)\, ,
\end{equation}

\noindent so that for the Bekenstein-Hawking entropy of the black hole 
we get, in units such that $G=\hbar =c=1$ 

\begin{equation}
S=\pi\left(2m^{2} +2mR_{0} -I_{2}\right)\, ,
\label{Entropy}
\end{equation}

\noindent where $I_{2}$ is the quadratic duality invariant defined 
in Eq.~(\ref{Invar1}). It is useful to have the expression of the entropy
in terms of the mass and supersymmetry central charges

\begin{equation}
S=\pi \left\{ (m^{2}-|{\cal Z}_{1}|^{2})
+ (m^{2}-|{\cal Z}_{2}|^{2})
+ 2\sqrt{(m^{2}-|{\cal Z}_{1}|^{2})(m^{2}-|{\cal Z}_{2}|^{2}) -J^{2}}\, .
\right\}\, .  
\end{equation}

For vanishing angular momentum $J=0$, this expression can be further
simplified to 

\begin{equation}
S=\pi \left[ \left(m^{2}-|{\cal Z}_{1}|^{2}\right)^{1/2}
+ \left(m^{2}-|{\cal Z}_{2}|^{2}\right)^{1/2}\right]^{2}\, ,
\end{equation}

\noindent which means that, if we believe the extrapolation of this 
formula to all extreme cases, the entropy vanishes if and only if both
Bogomol'nyi bounds are saturated and $1/2$ of the supersymmetries are
unbroken \cite{kn:KLOPP}.

When any one of the two possible Bogomol`nyi bounds is
saturated (for $J=0$) the entropy is proportional to the difference
between the modulus of the two central charges, which is proportional
to the quartic duality invariant $I_{4}$ defined in
Eq.~(\ref{Invar2}), which is moduli-independent. 

The temperature can be calculated imposing the regularity of the
metric near the event horizon in imaginary time. Following the
standard prescription~\cite{kn:GH}, we must shift the time
$t$ and the rotation parameter $\alpha$ to the values $t\rightarrow
i\tau$ and $\alpha\rightarrow i\tilde{\alpha}$ respectively. This
yields the Euclidean section of the metric, and the absence of conical
singularities at the event horizon in imaginary time requires the
identification $(\tau,\varphi) \sim (\tau+\beta_H,
\varphi-\tilde{\Omega}_H\beta_H)$, where $\tilde{\Omega}_H$ is the
Euclidean angular velocity of the event horizon and $\beta_H$ is the
inverse Hawking temperature. For the (real) angular velocity of the
horizon we have the following result:

\begin{equation}
\Omega_{H} = \frac{\alpha}{r_+^{\ 2}+\alpha^2-|\Upsilon|^2}\, ,
\end{equation}

\noindent and so we obtain, in a perfectly straightforward way, 
the value for the Hawking temperature of the black hole:

\begin{equation}
T_{H} =\frac{R_{0}}{2S}\, .
\label{Temperature}
\end{equation}

For $J=0$ the temperature always vanishes in the supersymmetric limit,
except in the case in which $1/2$ of the supersymmetries are going to
be left unbroken. In that case the limit is simply not well defined.


\section{Conclusions}
\label{sec-conclusions}

We have given a new set of solutions of pure $N=4$, $d=4$ supergravity
which are beyond the BPS limit (in both directions) and which
constitute the most general stationary point-like solution of this
theory, since all the conserved charges are present in our solution,
and all of them can take completely arbitrary values\footnote{The only
  possible addition would be primary scalar hair, but we are not
  interested in that kind of solutions.}. These solutions include
black holes as well as Taub-NUT spacetimes, BHs being non-extremal in
the general case. We have also shown that our family of solutions, and
also the thermodynamic quantities associated to the BHs, are
duality-invariant.

From a more technical point of view, we hope that the Ansatz providing
the solution (basically characterized by the introduction of the
`non-extremality' function $W$ and the non-flat three-dimensional
metric (\ref{Non-Flat}) as ``background'' space) will prove helpful
for the task of finding more non-extreme black holes s in other
models, in particular, in those arising from more realistic
compactifications of string theory, like compactifications on
Calabi-Yau spaces, orbifolds, etc.


\section*{Acknowledgments}

The work of E.L.-T.~is supported by a U.A.M.~grant for postgraduate
studies. The work of T.O.~is supported by the European Union TMR
program FMRX-CT96-0012 {\sl Integrability, Non-perturbative Effects,
  and Symmetry in Quantum Field Theory} and by the Spanish grant
AEN96-1655.

\appendix

\section{Conserved Charges and Duality Invariants}
\label{sec-conserved}

The non-geometrical conserved charges of this theory are associated to
the $U(1)$ vector fields. There are electric charges $\tilde{q}^{(n)}$
whose conservation law is associated to the Maxwell equation and are
defined, for point-like objects by the asymptotic behavior of the
$t-r$ components of the $SL(2,\mathbb{R})$-dual field strengths

\begin{equation}
\label{Cons-Ch1}
{}^{\star}\tilde{F}^{(n)}{}_{tr}
\sim  \frac{\tilde{q}^{(n)}}{\rho^{2}}\, ,      
\end{equation}

\noindent and magnetic charges $\tilde{p}^{(n)}$ whose conservation law 
is associated to the Bianchi identity and are defined, for point-like
objects by the asymptotic behavior of the $t-r$ components of the
field strengths

\begin{equation}
\label{Cons-Ch2}
{}^{\star}F^{(n)}{}_{tr}
\sim  \frac{\tilde{p}^{(n)}}{\rho^{2}}\, .
\end{equation}

These charges can be arranged in $SO(N)$ vectors $\vec{\tilde{q}},
\vec{\tilde{p}}$ and these can be arranged in $SL(2,\mathbb{R})$
doublets

\begin{equation}
\left(
\begin{array}{c}
\vec{\tilde{q}} \\
\\
\vec{\tilde{p}} \\
\end{array}
\right)\, .
\end{equation}

Under S~and T~duality transformations $\Lambda,R$, this charge vector
transforms according to

\begin{equation}
\left(
\begin{array}{c}
\vec{\tilde{q}}^{\, \prime} \\
\\
\vec{\tilde{p}}^{\, \prime} \\
\end{array}
\right)
=
\Lambda\otimes R
\left(
\begin{array}{c}
\vec{\tilde{q}} \\
\\
\vec{\tilde{p}} \\
\end{array}
\right)\, .
\end{equation}

It is also useful to introduce the $SL(2,\mathbb{R})$ matrix of scalar
fields

\begin{equation}
{\cal M} = e^{2\phi}
\left(
\begin{array}{cc}
|\lambda |^{2} & a \\
& \\
a              & 1 \\
\end{array}
\right)\, ,
\hspace{1cm}
{\cal M}^{-1} = e^{2\phi}
\left(
\begin{array}{cc}
1  & -a             \\
& \\
-a & |\lambda |^{2} \\
\end{array}
\right)\, ,
\end{equation}

\noindent which transforms under $SL(2,\mathbb{R})$ according to

\begin{equation}
{\cal M}^{\prime} = \Lambda {\cal M} \Lambda^{T}\, ,
\end{equation}

\noindent and it is an $SO(N)$ singlet.

We can now construct two expressions which are manifestly
$SL(2,\mathbb{R})\otimes SO(N)$-invariant and that will be useful
later to express physical results in a manifestly duality-invariant
way. The first one is quadratic in the charges:
  
\begin{equation}
\label{Invar1}
I_{2}\equiv
\left(
\begin{array}{cc}
\vec{\tilde{q}}\ \ \ \ 
\vec{\tilde{p}} \\
\end{array}
\right)
{\cal M}_{0}^{-1}
\left(
\begin{array}{c}
\vec{\tilde{q}} \\
\\
\vec{\tilde{p}} \\
\end{array}
\right)
\end{equation}

\noindent where ${\cal M}_{0}$ is the constant asymptotic value
of ${\cal M}$.

The second invariant we will use is :

\begin{equation}
  \label{Invar2}
I_{4}\equiv
\det \left[
\left( \begin{array}{c}
\vec{\tilde{q}} \\
\\
\vec{\tilde{p}} \\
\end{array} \right)
\left( \begin{array}{cc}
\vec{\tilde{q}} \ \ \ \
\vec{\tilde{p}} \\
\end{array} \right)
\right]=
\left(\vec{\tilde{q}}\cdot\vec{\tilde{q}}\right)^{2}
\left(\vec{\tilde{p}}\cdot\vec{\tilde{p}}\right)^{2}
-\left(\vec{\tilde{q}}\cdot\vec{\tilde{p}}\right)^{2}\, ,
\end{equation}

\noindent which is quartic in the charges. Observe that $I_{2}$ is 
moduli-dependent and $I_{4}$ is moduli-independent.


\section{Physical Parameters}
\label{sec-parameters}

Here we explain our notation for charges and moduli used in the
solutions. $m$ stands for the ADM mass, and $n$ for the NUT charge.
They appear combined into the complex constant ${\mathfrak{M}}$ defined by

\begin{equation}
\label{Mass}
{\mathfrak{M}} = m+in\, .
\end{equation}

The rotation parameter $\alpha$ is $\alpha=J/m$, where $J$ is the
angular momentum.

These charges are singlets under all duality transformations.

The asymptotic behavior of the axidilaton is characterized by the
constant asymptotic value $\lambda_{0}=a_{0} +ie^{-2\phi_{0}}$, where
$a_{0}$ is the constant asymptotic value of the axion and $\phi_{0}$
that of the dilaton. The axidilaton ``charge'' is denoted by
$\Upsilon$ and, thus

\begin{equation}
\lambda \sim \lambda_{0}-ie^{-2\phi_{0}} \frac{2\Upsilon}{\rho}\, .
\end{equation}

\noindent (where $\rho$ is a radial coordinate).

$\lambda_{0}$ transforms as $\lambda$ under duality transformations
and $\Upsilon$ is as $SO(N)$ singlet and, under $SL(2,\mathbb{R})$

\begin{equation}
\Upsilon^{\prime} = e^{-2i\arg(c\lambda_{0}+d)}\Upsilon\, .
\end{equation}

We find it convenient to use, instead of the conserved charges
$\tilde{q}^{(n)}$ and $\tilde{p}^{(n)}$ defined in
Appendix~\ref{sec-conserved}, the constants $Q^{(n)}$ and $P^{(n)}$
defined by

\begin{equation}
\label{E-M-Charges}
F_{tr}{}^{(n)}\sim \frac{e^{\phi_{0}}Q^{(n)}}{\rho^{2}}\, ,
\hspace{1cm}
{}^{\star}F_{tr}{}^{(n)}\sim -\frac{e^{\phi_{0}}P^{(n)}}{\rho^{2}}\, ,
\end{equation}

\noindent and combined into the complex constants $\Gamma^{(n)}$
which can be arranged into an $SO(N)$ vector

\begin{equation}
\Gamma^{(n)} = Q^{(n)}+iP^{(n)}\, ,
\hspace{1cm}
\vec{\Gamma} = \vec{Q}+i\vec{P}\, .
 \label{Dyonic}
\end{equation}

In our solutions these charges are simple combinations of the
conserved charges and moduli:

\begin{equation}
\left(
\begin{array}{c}
\vec{Q} \\
\\
\vec{P} \\
\end{array}
\right)  
=
{\cal V}_{0}
\left( 
\begin{array}{c}
\vec{\tilde{q}} \\
\\
\vec{\tilde{p}} \\
\end{array} 
\right)\, ,
\end{equation}

\noindent where

\begin{equation}
{\cal V}_{0} = 
e^{\phi_{0}}
\left( 
\begin{array}{cc}
-1 & a_{0}           \\
& \\
0  & -e^{-2\phi_{0}} \\
\end{array}
\right)\, ,
\hspace{1cm}
{\cal V}_{0}^{T} {\cal V}_{0} = {\cal M}_{0}^{-1}\, .
\end{equation}

$\vec{\Gamma}$ is an $SO(N)$ vector and transforms
under $SL(2,\mathbb{R})$ according to

\begin{equation}
\vec{\Gamma}^{\, \prime} = e^{i\arg(c\lambda_{0}+d)} \vec{\Gamma}\, ,
\end{equation}

\noindent so the duality invariants can be written

\begin{equation}
\begin{array}{rcl}
I_{2} & = & |\vec{\Gamma}\cdot\vec{\Gamma}|^{2} 
=\sum_{n=1}^{N}|\Gamma^{(n)}|^{2}\, ,\\
& & \\
I_{4} & = & -\frac{1}{4}{\rm det}
\left[
\left(
\begin{array}{c}
\vec{\Gamma}\\
\\
\bar{\vec{\Gamma}}\\
\end{array}
\right)
\left(
\begin{array}{cc}
\vec{\Gamma}& \ \ \
\bar{\vec{\Gamma}}\\
\end{array}
\right)
\right]\, .
\end{array}
\end{equation}

Our solutions do not have any primary scalar hair and the axidilaton
charge is always completely determined by the electric and magnetic
charges, and mass and NUT charge through

\begin{equation} 
\Upsilon =  -{\textstyle\frac{1}{2}}
\sum_{n=1}^{N}\frac{ (\bar{\Gamma}^{(n)})^{2}}{\mathfrak{M}}\, .
\label{Upsilon} 
\end{equation} 

The absolute value of this expression is duality invariant and can be
rewritten in terms of the basic invariants
(\ref{Invar1},\ref{Invar2}) as follows:

\begin{equation}
|\Upsilon|^{2} = \frac{1}{4|\mathfrak{M}|^{2}} (I_{2}^{2}-4I_{4})\, .
\end{equation}


\section{Central Charge Matrix Eigenvalues}
\label{sec-centralcharge}

The supersymmetry parameter $r_{0}$ can be expressed in terms of the
two different skew eigenvalues of the central charge matrix of
$N=4,d=4$ Supergravity~\cite{kn:WO, kn:FSZ} ${\cal Z}_{1,2}$. Their
absolute values can be expressed in terms of the electric and magnetic
charges in the following way:

\begin{equation}
|{\cal Z}_{1,2}|^{2}=
{\textstyle\frac{1}{2}}
\displaystyle{\sum_{n=1}^{N}}|\Gamma^{(n)}|^{2}\pm
{\textstyle\frac{1}{2}}
\left[ 
\left( 
\displaystyle{\sum_{n=1}^{N}}|\Gamma^{(n)}|^{2}
\right)^{2}
-\left|
\displaystyle{\sum_{n=1}^{N}}\left(\Gamma^{(n)}\right)^{2}
\right|^{2}
\right]^{1/2}\, ,
\end{equation}

\noindent and in terms of the invariants $I_{2},I_{4}$
defined in Eqs.~(\ref{Invar1},\ref{Invar2}) as follows

\begin{equation}
|{\cal Z}_{1,2}|^{2}=
{\textstyle\frac{1}{2}}I_{2} \pm I_{4}^{1/2}\, .  
\end{equation}

With the help of these expressions and those of the previous
Appendices we can write the supersymmetry parameter $r_{0}$ as
follows:

\begin{equation}
r_{0}^{2}=
\frac{1}{|{\mathfrak{M}}|^2}\left(|{\mathfrak{M}}|^2-|{\cal Z}_1|^2 \right)
\left(|{\mathfrak{M}}|^2-|{\cal Z}_2|^2 \right)\, .
\label{r0-Zs}
\end{equation}

This last equation makes explicit the fact that, if and only if
$r_0^{2}$=0, one of the two possible supersymmetry Bogomol'nyi bounds

\begin{equation} 
|{\mathfrak{M}}|^{2} \geq |{\cal Z}_{1,2}|^{2}\, , 
\end{equation}

\noindent is  saturated.

The following expression is also useful

\begin{equation}
|{\cal Z}_{1}{\cal Z}_{2}|^{2} = |\mathfrak{M}|^{2} |\Upsilon|^{2}\, .  
\end{equation}


\end{document}